\documentclass[%
prl,
twocolumn,
nofootinbib,
nobibnotes,
amsmath,amssymb,
aps,
]{revtex4-1}

\usepackage{graphicx}
\usepackage{dcolumn}
\usepackage{bm}
\usepackage{hyperref}
\usepackage{graphicx}
\usepackage[usenames, dvipsnames]{color}

\usepackage[utf8]{inputenc}
\usepackage[normalem]{ulem}

\begin{document}
	
	
\title{Numerical Studies on Core Collapse Supernova in Self-interacting Massive Scalar-Tensor Gravity}
	
\author{Patrick Chi-Kit \surname{Cheong}}
\email{chi-kit.cheong@ligo.org}
\affiliation{%
	Department of Physics, The Chinese University of Hong Kong, Shatin, N. T., Hong Kong
}%
\author{Tjonnie Guang Feng \surname{Li}}%
\email{tgfli@cuhk.edu.hk}
\affiliation{%
	Department of Physics, The Chinese University of Hong Kong, Shatin, N. T., Hong Kong
}%
	
\date{\today}

\begin{abstract}
We investigate stellar core collapse in scalar-tensor theory with a massive self-interacting scalar field.
In these theories, strong long-lived inverse chirp signals could be induced during the stellar core-collapse which provides us with several potential smoking-gun signatures that could be found using current ground-based detectors.
We show that the existence of self-interaction in the potential of the scalar field can significantly suppress spontaneous scalarization and the amplitude of the monopole gravitational wave radiation.
Moreover, this suppression due to self interaction is frequency dependent and may be discernible in LIGO/Virgo's sensitive band.
Therefore, self-interaction should be considered when constraining scalar-tensor coupling parameters with gravitational-wave detections.
Alternatively, if such monopole gravitational-wave signals are detected, then one may be able to infer the presence of self-interaction.
	\end{abstract}
	
\pacs{Valid PACS appear here}
\maketitle


\section{\label{sec:intro}Introduction}
General Relativity (GR) has passed a large number of tests which scale from submillimeter-scale to the Solar System scale and in astrophysical and cosmological scales (see e.g. \cite{Will2014, Psaltis2008,2015CQGra..32x3001B} and references therein). 
The nonrenormalizability of GR in quantum field theory suggests that GR should be regarded as an effective theory.
In addition, with the existence of dark matter and dark energy, modifications of GR may be unavoidable in different energy regimes.

Most of the tests of GR was performed in the relatively weak field regime.
By means of gravitational waves observations by LIGO-Virgo \cite{PhysRevLett.116.061102}, we are able to probe the extreme physics in the strong-field gravity and it provides us with another way to test Einstein's theory \cite{PhysRevD.94.084002,PhysRevLett.116.221101}.
One way to test GR based on GW detections is to choose an alternative theory and see rather it can explain the data well or not.

A possible cosmological and astrophysical extension of GR is the scalar-tensor (ST) theory of gravity \cite{STT,fujii_maeda_2003}.
This theory has been studied over the past decades and the mathematical understanding of scalar-tensor theories has matured enough allow for fully nonlinear numerical simulations \cite{2014PhRvD..89h4005S, 2015CQGra..32t4001H}. 
Scalar-tensor theories allow for GWs in the well-tested weak-field regime, yet show significant deviations in strong gravity. 
For example, the existence of nonperturbative strong-field effects in neutron stars, also known as \emph{spontaneous scalarization} \cite{Scalarization}.

The parameter space of scalar-tensor theories is considerably constrained through various astrophysical observations, for example, by measuring the orbital decay of binary pulsars \cite{binary_test1, binary_test2}.
No significant GR deviations are allowed within the narrow parameter space of these theories.
However, the parameter space of scalar-tensor theories is weakly constrained if we extend the theory with a massive scalar field.
For instance, for a scalar field with mass $\mu$, the corresponding Compton wave-length is $\lambda_{\varphi} = 2\pi\hbar/(\mu c)$.
The observations mentioned above cannot be applied if the length scale of those systems is greater than the Compton wavelength of the scalar field $\lambda_{\varphi}$.
Therefore, the bounds from those observations are only valid on extremely light (i.e. the scalar mass $\mu \lesssim 10^{-19}$ eV) or even massless scalar-tensor theories \cite{scalarization_massive_1,  scalarization_massive_2}.

Spontaneous scalarization in neutron stars with massive scalar fields has been well studied in static, slowly rotating and rapidly rotating cases \cite{2016PhRvD..93f4005R, 1475-7516-2016-11-019,2017PhRvD..96h4026M}. 
All of those studies show that the observationally allowed range of the scalar-tensor parameters is dramatically changed if the mass of the scalar field was taken in the count.
The studies on 
neutron stars with the natural extension with self-interaction was just recently initiated \cite{self_interact}.

An interesting channel to study this class of theories is therefore in the formation of neutron stars.
Massive star with zero-age main sequence (ZAMS) masses in the range of $8M_{\odot} \lesssim M_{\text{ZAMS}} \lesssim 100 M_{\odot}$ die as a core- collapse supernova (CCSN).
A proto-neutron star is formed during the whole dynamical process, and is left as a neutron star if the CCSN explodes successfully or otherwise becomes a black hole.
Therefore, core-collapse supernovae are the testbed for us to investigate the spontaneous scalarization dynamically as it forms neutron stars and black holes. 

Not only for the stationary cases, but the spontaneous scalarization has also been studied in the fully non-linear dynamical regime.
A recent study shows that in massive scalar-tensor theory, hyperscalarization could be induced during the stellar core collapse.
It also generates dispersive hyperscalarized long-lived inverse chirp monopole GW signals \cite{PhysRevLett.119.201103}.
The signal depends mainly on a coupling parameter for the scalar-tensor theory and the amplitude of the scalar field.
As the simulations suggest that the amplitude of the scalar field is intrinsic and insensitive to many parameters, one can put an impressive constraint on scalar-tensor parameters by assuming no such detection.

Inspired by the studies above, we present the initial numerical study of dynamical strong fields in stellar core collapses in scalar-tensor theory with a \emph{massive self-interacting} scalar field by investigating the scalar GW signature and explore the scalar-tensor parameter space.

The paper is organised as follows.
In Sec.~II we outline the formalism we used in this work.
The details of the numerical simulation settings and results are presented in Sec.~III.
This paper ends with a discussion section in Sec.~IV.

\section{\label{sec:methods}Methods}
The scalar-tensor theory action in Einstein frame is given by (using natural units $G=1=c$)
\begin{eqnarray}
	\nonumber S =& \frac{1}{16 \pi}\int dx^4 \sqrt{-\bar{g}} \left[\bar{R} -2\bar{g}^{\mu \nu}(\partial_\mu \varphi)(\partial_\nu \varphi)-4V(\varphi) \right] \\ 
	&+S_m(\psi_m,g_{\mu \nu}/F),
	\label{eq:action}
\end{eqnarray}
where $\bar{R}$ is the Ricci scalar, $\varphi$ and $V(\varphi)$ are the scalar field and the potential respectively. 
Note that barred variables are constructed from the conformal metric $\bar{g}_{\mu\nu}={g}_{\mu\nu}/F(\varphi)$, where $g_{\mu\nu}$ is the physical or Jordan-Fierz metric and $F(\varphi)$ is the coupling function. 
Once the conformal factor $F(\varphi)$ and the potential $V(\varphi)$ are chosen, the theory is specified. 
In this work, we study scalar-tensor theories with quadratic coupling functions, which are widely used in the literature \cite{Scalarization,damour1996tensor}
\begin{equation}
	F(\varphi) = \exp(-2\alpha_0\varphi-\beta_0\varphi^2),
	\label{eq:coupling}
\end{equation}
where $\alpha_0$ and $\beta_0$ are two free parameters in the coupling function $F(\varphi)$.
Moreover, we use the potential of a massive scalar field with a quartic self-interaction \cite{PhysRevLett.119.201103, self_interact}
\begin{equation}
	V(\varphi) = \frac{\mu^2}{\hbar^2}\frac{\varphi^2}{2} + \lambda \varphi^4, 
	\label{eq:potential}
\end{equation}
where $\mu$ is the mass of the scalar field and $\lambda$ is a non-negative coupling constant.

We follow the equations of motion from Refs.~\cite{PhysRevLett.119.201103, massless_ST} and implemented them to \texttt{GR1D} \cite{GR1D}.
\texttt{GR1D} is a open-source spherically-symmetric general relativistic hydrodynamics code for stellar collapse to neutron stars and black holes.
It is able to capture many qualitative aspects of CCSNe and was used to study different stellar collapse and BH formation scenarios \cite{2010AIPC.1269..166O}.
As in \cite{PhysRevLett.119.201103}, we assume spherical symmetry, use high-resolution shock capturing scheme for matter evolution and do the simulation with a phenomenological hybrid equation of state (EOS).
All equations, discretization, grid and boundary treatment are identical to Ref.~\cite{PhysRevLett.119.201103}, except for the potential with a self-interacting term.

\section{\label{sec:results}Simulations and Results}
The simulations are specified by seven parameters: the mass of the scalar field $\mu$, a coupling constant $\lambda$ for the self-interaction, two parameters $\alpha_0$ and $\beta_0$ for the coupling function $F(\varphi)$, two adiabatic indices $\Gamma_1$, $\Gamma_2$ for subnuclear, supranuclear polytropic EOS and the thermal adiabatic index $\Gamma_{\text{th}}$ for the thermal part pressure which models a mixture of relativistic and non-relativistic gas. 
Similar to \cite{PhysRevLett.119.201103}, we used realistic non-rotating pre-SN models WH12 from \cite{WOOSLEY2007269} as our initial profile with initially vanishing scalar field and varying the parameters in the ranges as shown in the table \ref{table:ranges}.
\begin{table}
	\caption{\label{tab:example} Ranges and values of the scalar-tensor and EOS parameters (following Ref.~\cite{PhysRevLett.119.201103}) explored in our 1-dimensional core-collapse supernova simulations.} 
\begin{ruledtabular}
\begin{tabular}{lll}
	Parameter & & Range/Value \\ \hline
	Coupling function parameter & $\alpha_0$ & $[10^{-4},10^{-2}]$  \\ 
	Coupling function parameter & $\beta_0$ & $[-5,-20]$  \\ 
	Scalar field mass & $\mu$ (eV) & $[0,10^{-13}]$  \\ 
	Self-interaction term & $\lambda$ & $[10^{-11},1]$  \\ 
	First adiabatic index & $\Gamma_{1}$ & $1.3$  \\ 
	Second adiabatic index& $\Gamma_{2}$ & $\{2.5,3\}$  \\ 
	Thermal adiabatic index& $\Gamma_{\text{th}}$ & $\{1.35,1.5\}$  \\ 
\end{tabular}
\end{ruledtabular}
\label{table:ranges}
\end{table}

\begin{figure}
	\centering
	\includegraphics[width=\columnwidth, angle=0]{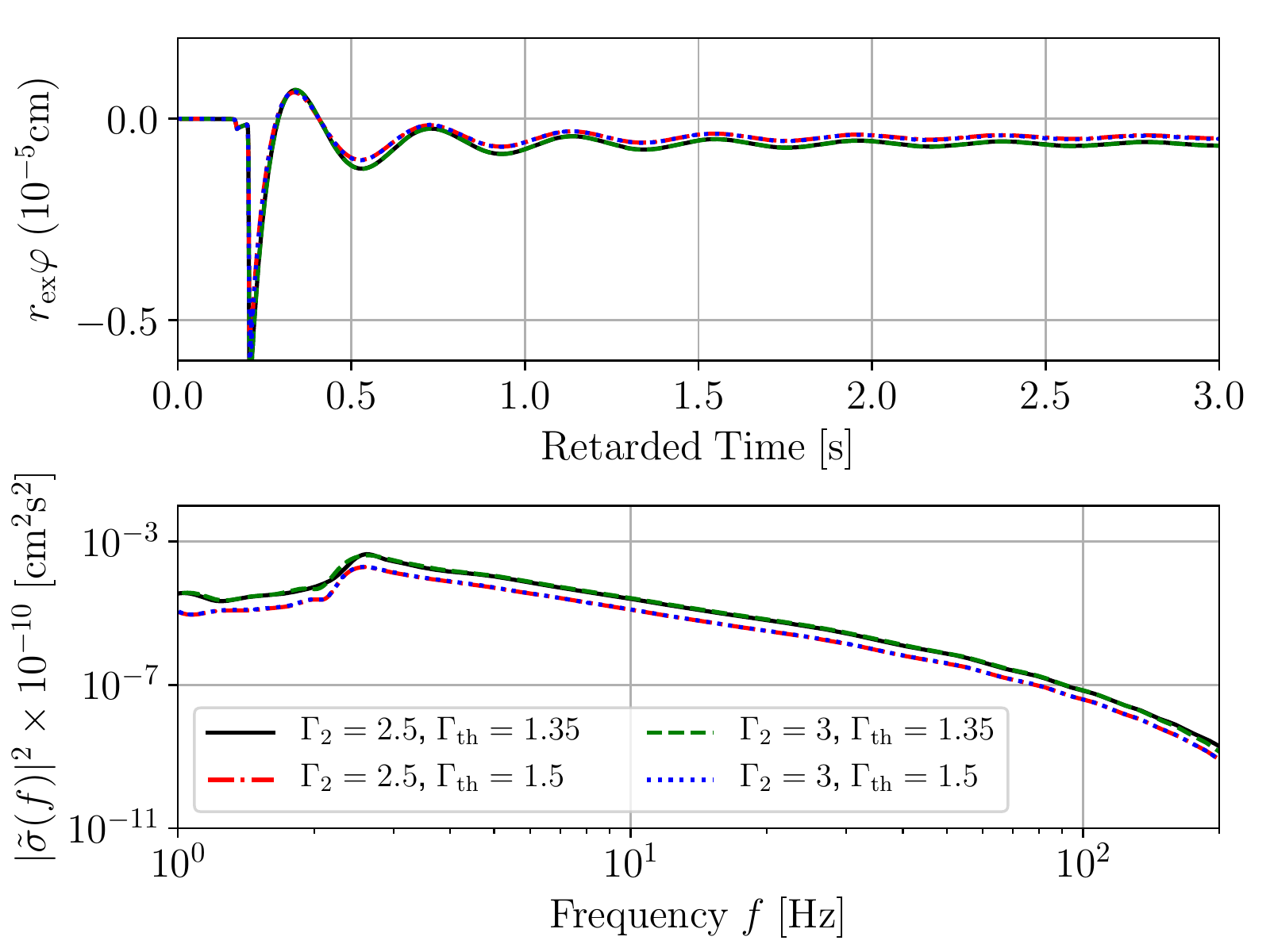}
	\caption{
		Upper panel: The monopole GWs extracted at $5\times10^4$ km with different EOS parameters.
		Lower panel: The power spectral density of the signal with different EOS parameters. 
	The frequency distributions of the scalar field $\varphi$ are insensitive to the EOS parameters. }
	\label{fig:EOS}	
\end{figure}

The EOS could significantly affect the dynamics of the core collapse supernovae\cite{PhysRevD.78.064056,PhysRevLett.98.251101}. 
	However, it was shown that for massive-scalar theories, the scalar field is insensitive to the detail of the source if the ST parameter $\beta_0$ is sufficiently negative \cite{PhysRevLett.119.201103}.
	We further demonstrate that this is still valid even with the self-interaction parameter $\lambda$ included.
To see if the effect due to different EOS is imprinted in the monopole GWs from the stellar core collapse, we first study the EOS parameters with other parameters fixed (i.e. $\mu = 10^{-14}$ eV, $\alpha_0 = 10^{-2}$, $\beta_0=-20$, $\lambda=10^{-1}$). 
Fig. \ref{fig:EOS} shows the power spectral density of the signal with different EOS parameters. 
We find that the frequency distributions of the scalar field $\varphi$ are insensitive to the EOS parameters.
In order to focus on the effects of self-interaction, we present the results of simulations with the typical values of $\Gamma$s \cite{GR1D} and set a sufficiently negative $\beta_0$ which induces the strong scalarization for $\lambda = 0$ cases \cite{PhysRevLett.119.201103}.
Specifically, in the following studies, we perform the simulation with $\mu = 10^{-14}$ eV, $\alpha_0 = 10^{-2}$, $\beta_0=-20$, $\Gamma_1 = 1.3$, $\Gamma_2 = 2.5$, $\Gamma_{\text{th}} = 1.35$ as in \cite{PhysRevLett.119.201103} as our reference parameters set and with the self-interaction parameter $\lambda$ which we vary in range $10^{-11} \leq \lambda \leq 1$. 

\subsection{Suppression of spontaneous scalarization}
\begin{figure}
	\centering
	\includegraphics[width=\columnwidth, angle=0]{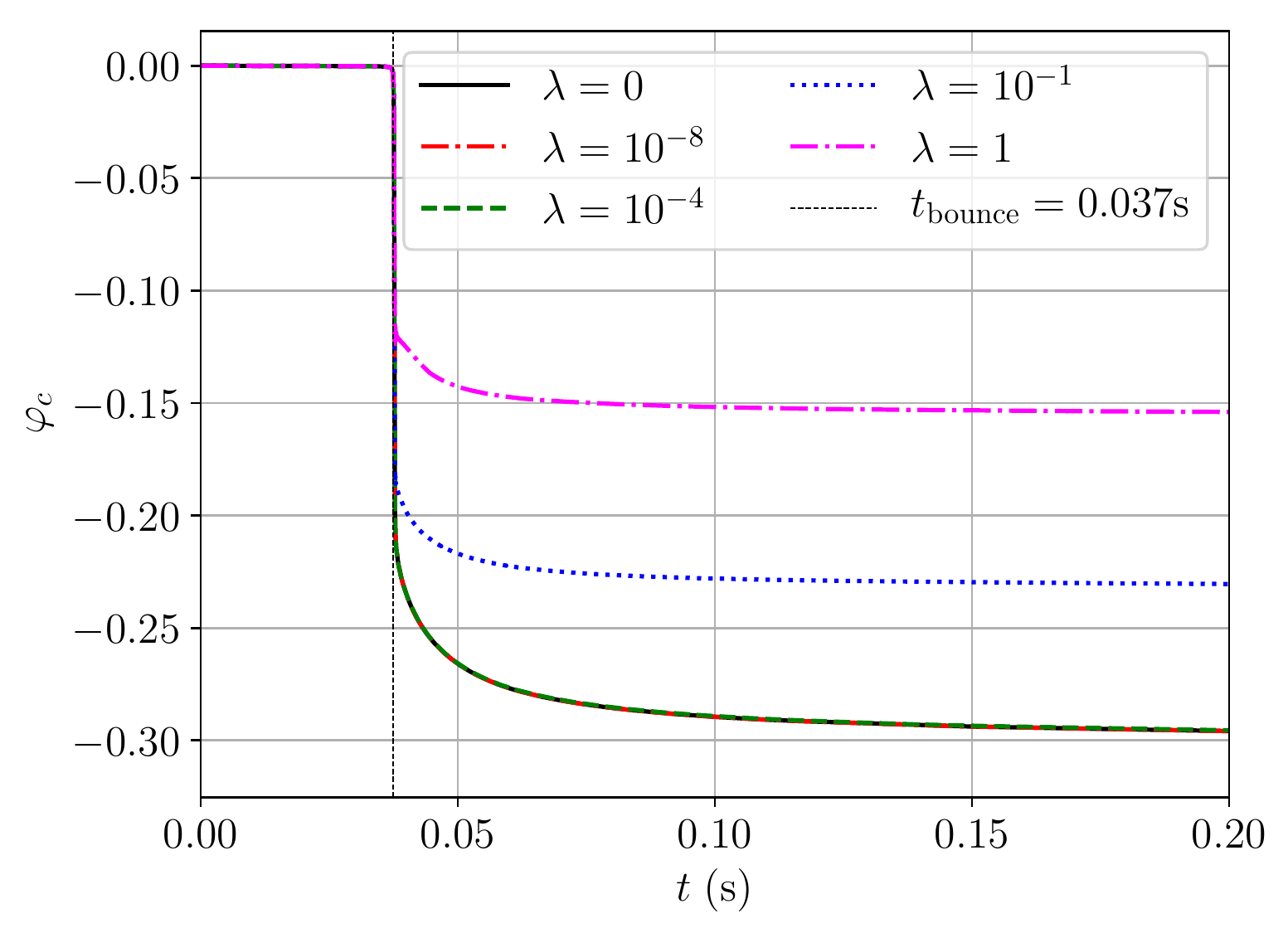}
	\caption{
	The scalar field $\varphi(r,t)$ at the center of the star as the function of time (i.e. $\varphi_c(t) = \varphi(r=0, t)$).
	The plots show that spontaneous scalarization arises at the core bounce for all values of the self-interaction parameter $\lambda$.
	The existence of self-interaction of the scalar field progressively suppress the spontaneous scalarization. }
	\label{fig:phi_c_lambdas}
\end{figure}

\begin{figure}
	\centering
	\includegraphics[width=\columnwidth, angle=0]{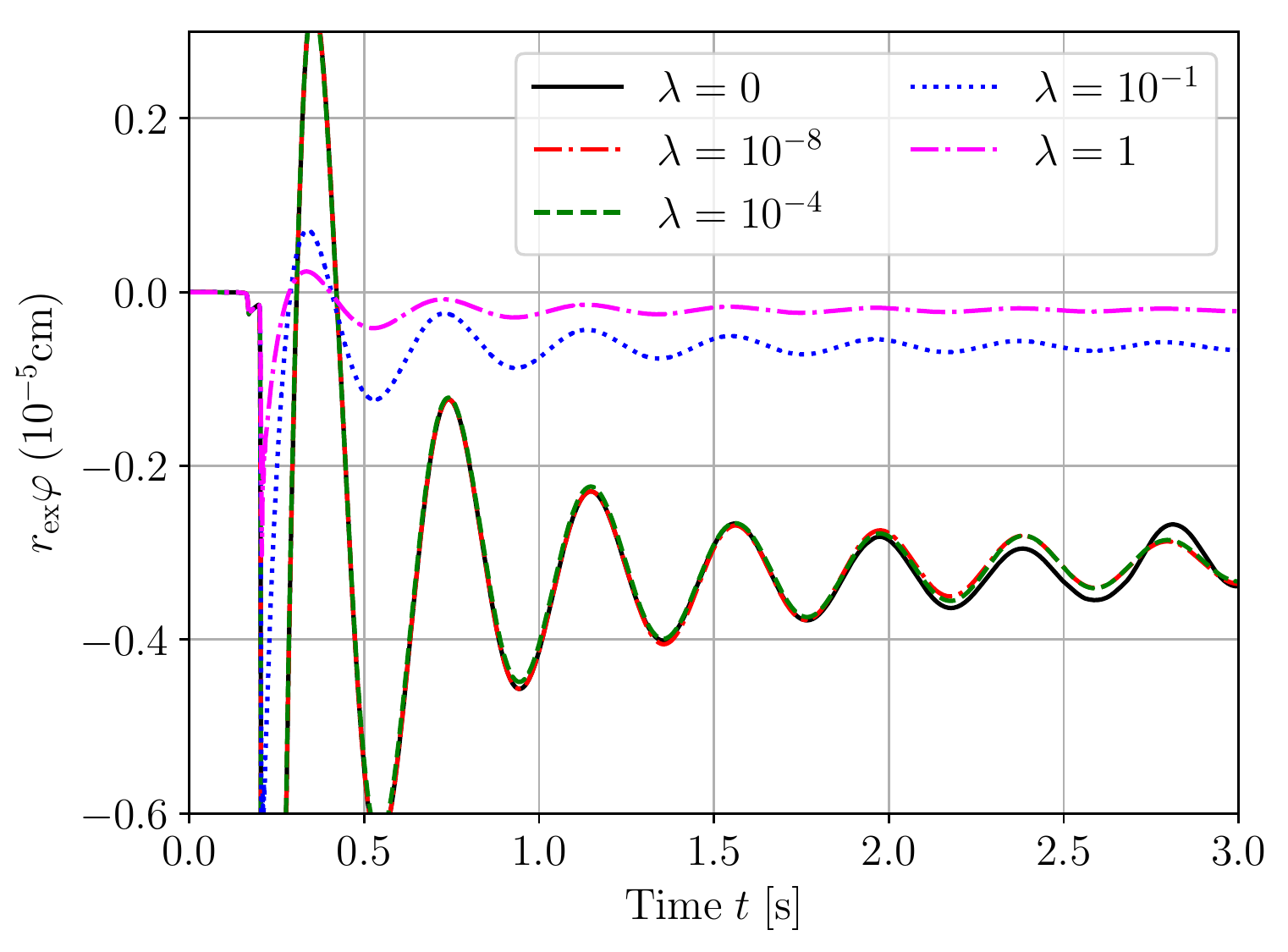}
	\caption{
	Waveforms $\sigma(r,t) \equiv r \varphi $ extracted at $5\times10^4$ km.
	The existence of self-interaction of the scalar field significantly reduce the amplitude and the memory effect of the scalar field. }
	\label{fig:lambdas}
\end{figure}

\begin{figure}
	\centering
	\includegraphics[width=\columnwidth, angle=0]{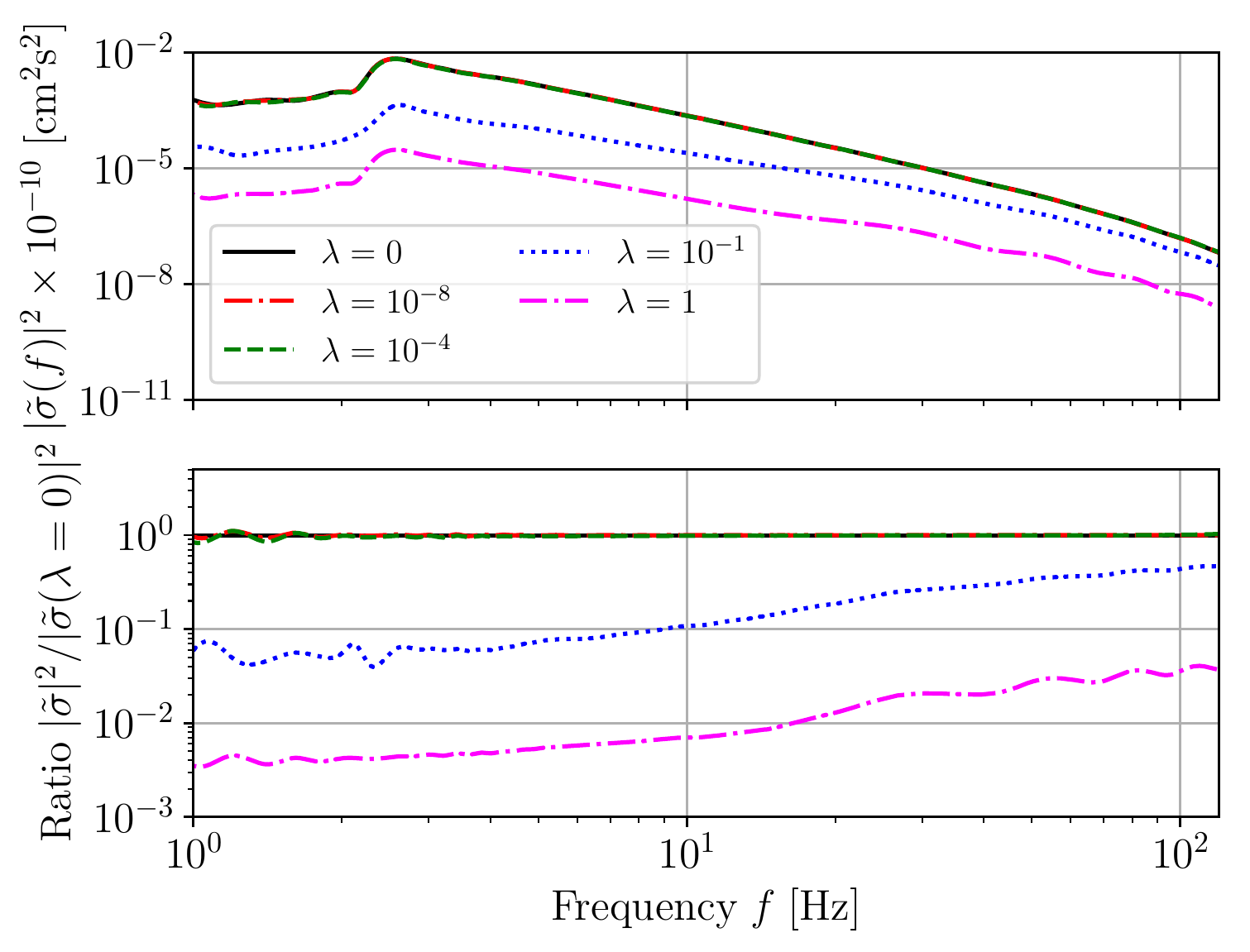}
	\caption{
		Upper panel: The power spectral density of the signal in the Fig. \ref{fig:lambdas} in frequency domain.
                Lower panel: The ratio of the PSD with or without self-interaction $\lambda$.
                Although the existence of the self-interaction suppresses the spontaneous scalarization and the scalar field, the frequency distributions of the signals in the low-frequency domain are almost the same.
                As shown in the lower panel, the suppression depends on the frequency.
                The low-frequency portion of the signal is suppressed significantly by self-interaction.
                The effect gradually decreases as the frequency increases.
		}
	\label{fig:PSD_lambdas}
\end{figure}

Fig. \ref{fig:phi_c_lambdas} shows the scalar field $\varphi(r,t)$ at the center of the star as the function of time (i.e. $\varphi_c(t) = \varphi(r=0, t)$). 
This figure shows that the spontaneous scalarization arises at the core bounce for all values of the self-interaction term $\lambda$.
However, the scalarization is suppressed progressively when the self-interaction term $\lambda$ is getting larger.
Even for sufficiently negative $\beta_0$, the hyperscalarization as mentioned in Ref.~\cite{PhysRevLett.119.201103} may not be that ``hyper'' if the self-interaction term $\lambda$ is not negligible.

Indeed, recent studies show that in cases of static and slowly rotating neutron stars, the scalarization is suppressed due to the self-interaction term \cite{self_interact}.
For fixed scalar-tensor parameters, the GR deviation decreases as the coupling constant $\lambda$ in the self-interaction increases, which is consistent with our observations.
We can understand this by noting that in the equation of motions of the scalar field, $\partial_t\psi$ is related to the $\partial_t^2\varphi$, which behaves like the ``driving force'' of the scalar field $\varphi$ \cite{massless_ST}.
The scalar potential induces an additional term $-\alpha F V_{,\varphi}$ in the $\partial_t\psi$ in Eq. (6) in ~\cite{PhysRevLett.119.201103}, where in our case $V_{,\varphi} = \left(\frac{\mu^2}{\hbar^2} + 4 \lambda \varphi^2 \right)\varphi$. 
With the negative $\varphi_c$, the existence of the self-interaction term $\lambda$ makes $\partial_t \psi$ less negative at $r=0$ and hence the scalarization is expected to be suppressed.

\subsection{Monopole gravitational-wave signals and propagation}
Besides the suppression of spontaneous scalarization, the monopole gravitational-wave signals are also suppressed by the self-interaction.
In Figs. \ref{fig:lambdas} and \ref{fig:PSD_lambdas}, we plot the gravitational-wave signal $\sigma \equiv r \varphi$ extracted at $5\times10^4$ km and the corresponding power spectral density in frequency domain with various coupling constant of the self-interaction $\lambda$.
Fig. \ref{fig:lambdas} shows that self-interaction could strongly suppress the amplitude of the signals and reduce the memory effect.
For instance, the power spectral density for the case with self-interaction term $\lambda = 1$ is $ \sim 1000$ smaller than the no self-interaction case ($\lambda = 0$)
Here we note that the units in Fig.~\ref{fig:phi_c_lambdas} and Fig.~\ref{fig:PSD_lambdas} are different.
In Fig.~\ref{fig:phi_c_lambdas}, we plot the scalar field at the centre directly.
While in Figs.~\ref{fig:lambdas} and \ref{fig:PSD_lambdas}, the waveforms are defined as $\sigma \equiv r \varphi$ and the power-spectral density as $|\sigma|^2$.
If the waveform is suppressed by a factor of 31.6, it would be suppressed roughly 1000 times in the power spectral density.
In conclusion, the existence of the self-interaction of the scalar field significantly affects the detectability of the monopole GW from the collapsing stars in massive self-interacting scalar-tensor theory.

Finally, the suppression of the signals is found to be frequency dependent. 
The lower panel in Fig.~\ref{fig:PSD_lambdas} shows the ratio of the power spectral density with or without self-interaction. 
In particular, the low-frequency portion of the gravitational-wave signal is suppressed significantly by the self-interaction, but this suppression gradually diminishes as the frequency increases. 
This frequency dependence may be explained as follows:
At large distances from the source, we can express the wave equation in the flat-spacetime approximately, namely
\begin{equation}
	\label{eq:wave}
\frac{\partial^2 \sigma}{\partial t^2} - \frac{\partial^2 \sigma}{\partial r^2} + \frac{\mu^2}{\hbar^2}\sigma + 4\lambda \frac{\sigma^3}{r^2} = 0\;,
\end{equation}
where $\sigma \equiv r \varphi$. 
The self-interaction is short range for the wave propagation due to the $1/r^2$ fall off.
Moreover, the plane waves propagate with group velocity $v_{g} \approx [1-(\omega_*^2/\omega^2)]^{1/2}$ for $\omega > \omega_* \equiv \mu/ \hbar$ \cite{PhysRevLett.119.201103}.
The group velocity is lower for the lower frequencies.
Therefore, the self-interaction is expected to have a larger effect in the lower frequencies as it takes longer for the low frequencies to propagate out of the ``self-interacting regime''. 
This frequency-related signature lies in the LIGO/Virgo sensitive band (i.e. $10\,{\rm Hz} - 10^3\,{\rm Hz}$).
Therefore, if such monopole gravitational waves are detected, then one may be able to infer if the scalar field is self-interacting.

\begin{figure}
	\centering
	\includegraphics[width=\columnwidth, angle=0]{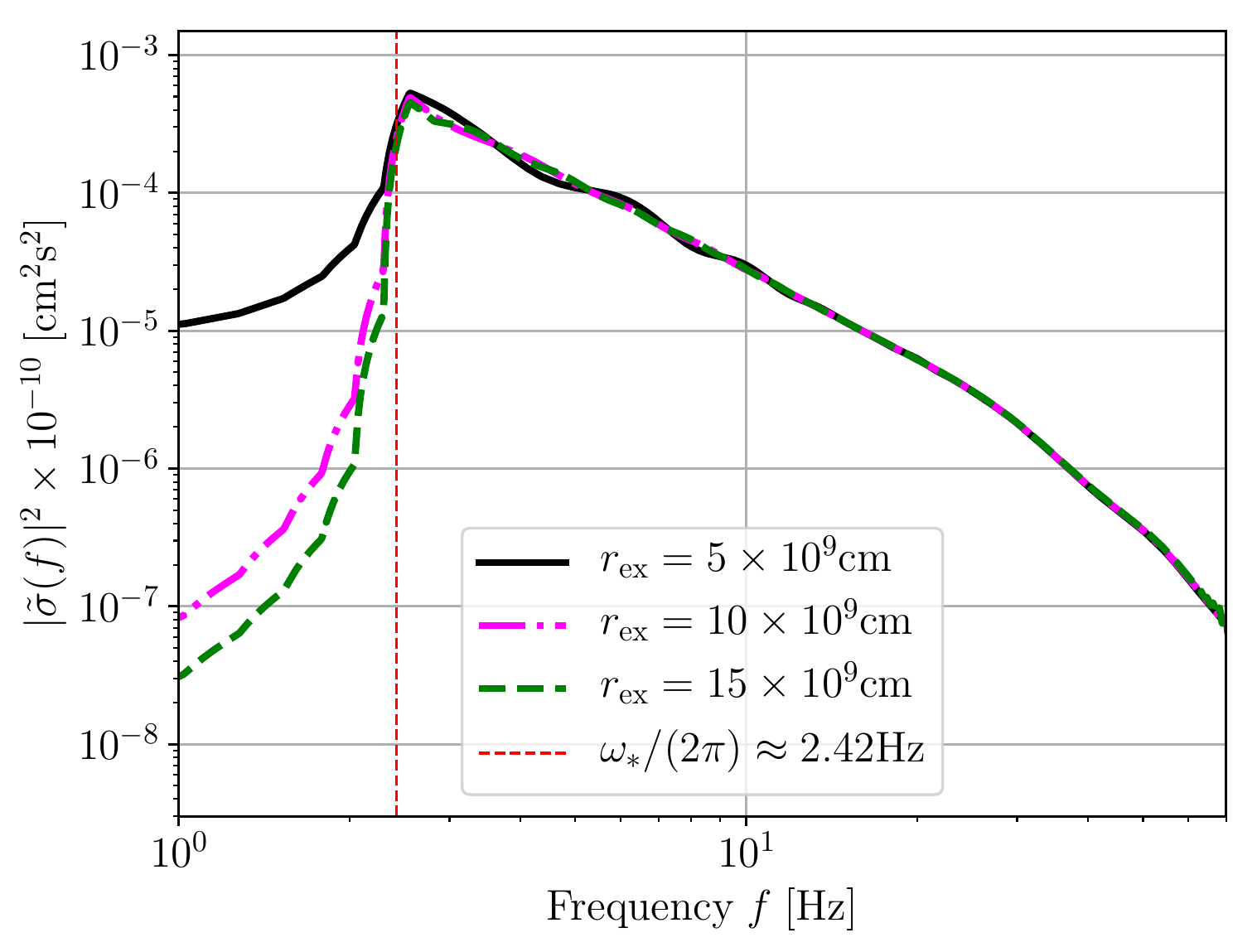}
	\caption{The power spectral density of the signal (with $\lambda=0.1$) at different distance. 
	The frequencies below $\omega_{*} \equiv \mu/ \hbar$ are damped exponentially but the part which above the critical frequency almost remains unaltered.
	}
	\label{fig:diss}
\end{figure}
The signal will not be the same as shown in Fig. \ref{fig:lambdas} at larger distances due to the dispersive nature of the scalar field.
In this case, the $\lambda$ term can be ignored since the distance between the observer and the source $r$ is normally on astrophysical scales.
The signal with the angular frequencies lower than $\omega_{*} \equiv \mu/ \hbar$ are damped exponentially which gives a similar conclusion as in \cite{PhysRevLett.119.201103}. 
This also suggest that if the scalar field is massless, the frequency dependence will become weaker.
To demonstrate this, we extracted the waveform at various distances as shown in Fig.~\ref{fig:diss}. 
As the signal propagates, the signal becomes increasingly oscillatory and the memory effect is suppressed significantly.
Frequencies below $\omega_{*}/(2\pi)$ are damped exponentially, but the part above the critical frequency remains unaltered.
Therefore, the stationary phase approximation is still valid and yields an inverse chirp-like signal that may result in a near monochromatic signal at the detector frame.
The three GW-search strategies (i.e. monochromatic, burst and stochastic searches) proposed in \cite{PhysRevLett.119.201103} remain applicable in the case of a self-interacting scalar field.

\section{\label{sec:discussion}Discussion}
We extended an open-source code \texttt{GR1D} \cite{GR1D} with massive self-interacting scalar-tensor gravity theories and performed the numerical simulations to study the strong-field dynamics and the monopole gravitational waves generation in stellar core collapse.

We find that the amplitude of the scalar field is insensitive to the EOS parameters, and depends weakly on the scalar-tensor coupling parameters $\alpha_0$, $\beta_0$ and the mass of the scalar field $\mu$, similar to Ref.~\cite{PhysRevLett.119.201103}.
The self-interaction parameter $\lambda$ suppresses the spontaneous scalarization and the whole evolution of the scalar field.
Even for sufficiently negative $\beta_0$, the scalarization is suppressed significantly if the coupling constant of the self-interaction $\lambda$ is large enough.

Moreover, we show that the dispersion relation of the scalar field at large distance of the source depends mainly on the mass of the scalar field $\mu$ and the effects due to the coupling constant of the self-interaction $\lambda$ can be ignored.
Therefore, we recover the results from Ref.~\cite{PhysRevLett.119.201103} that the scalar gravitational-wave signal disperses as it propagates through astrophysical distances and becomes an inverse chirp signal.

For a different range of the scalar-tensor parameters, the signature of the GW signal from stellar collapse in scalar-tensor theory may occur in continuous wave searches, stochastic searches and burst searches.
For all these types of searches, the effect due to the self-interaction should be considered.
In case of monochromatic searches, the signal can be described mainly by the magnitude of the scalarization (which affects the amplitude of the signal) and the mass $\mu$ of the scalar field.
Although the GW strain scales linearly with $\alpha_0$ (i.e. $h \propto \alpha_0 \varphi$), we cannot simply put a constraint on $\alpha_0$ by the non-detection of such signals as the magnitude of the scalarization could be suppressed significantly by the self-interaction of the scalar field.

Moreover, the amount of the self-interaction induced suppression of the signals depends on the frequency of the signals.
Since this frequency related signature lies in the LIGO/Virgo sensitive band, one may be able to infer the existence of self-interacting through gravitational-wave measurements.

It is well known that microphysics and neutrino physics play a large role in core-collapse supernovae, 
the dynamical features of core-collapse supernovae will no longer the same if we make use of realistic EOS and implement a proper neutrino treatment.
The matter evolution will significantly affect the evolution of the scalar field, the GWs signature should be changed dramatically.
With more detailed and realistic input physics, not only we are able to study how those inputted physics imprinted in the monopole GWs, but also the resulting neutrino luminosity might provide us with another way to constraint the scalar-tensor theories.
This will be left for future work.

\section{Acknowledgments}
We thank U.~Sperhake for giving us the pointers when we started the project, and L.~M.~Lin for detailed discussions and suggestions on numerical methods.
This work was partially supported by grants from the Research Grants Council of the Hong Kong (Project No. CUHK 14310816 and CUHK 24304317) and by the Direct Grant for Research from the Research Committee of the Chinese University of Hong Kong.

\bibliographystyle{apsrev4-1}
\bibliography{cite}

\end{document}